\documentclass[12pt]{iopart}

\usepackage{graphicx}
\usepackage{setspace}

\begin{document}

\title[ ]{Rashba-Zeeman-effect-induced spin filtering energy windows in a quantum wire}

\author{Xianbo Xiao$^{1}$\footnote{E-mail: xxb\_11@hotmail.com}, Zhaoxia Chen$^{2}$, Wenjie Nie$^{1}$, Guanghui Zhou$^{3}$, and Fei Li$^{4}$\footnote{E-mail: wltlifei@sina.com} }

\address{$^1$ School of Computer, Jiangxi University of Traditional Chinese Medicine, Nanchang 330004, China.\\
$^2$ School of Mechatronics Engineering, East China Jiaotong University, Nanchang 330013, China.\\
$^3$ Department of Physics and Key Laboratory for Low-Dimensional Quantum Structures and Manipulation (Ministry of Education), Hunan Normal University, Changsha 410081, China.\\
$^4$ Office of Scientific Research, Jiangxi University of Traditional Chinese Medicine, Nanchang 330004, China.}

\begin{abstract}
We perform a numerical study on the spin-resolved transport in
a quantum wire (QW) under the modulation of both Rashba spin-orbit coupling
(SOC) and a perpendicular magnetic field (MF) by adopting the developed Usuki
transfer-matrix method in combination with the Landauer-B$\mathrm{\ddot{u}}$ttiker formalism.
Wide spin filtering energy windows can be achieved in
this system for a spin-unpolarized injection. In addition, both the width
of these energy windows and the magnitude of the spin conductance within
these energy widows can be tuned by varying the Rashba SOC strength, which
can be apprehended by analyzing the energy dispersions and the
spin-polarized density distributions inside the QW, respectively. Further study
also demonstrates that these Rashba-SOC-controlled spin filtering energy windows
show a strong robustness against disorders. These findings may not
only benefit to further understand the spin-dependent transport properties of the
QW in the presence of external fields but also provide a theoretical instruction
to design a spin filter device.

\end{abstract}

\maketitle

\section{Introduction}

The charge transport properties of quantum wires (QWs) formed in a two-dimensional
electron gas (2DEG) by a split gate technique \cite{Wees} have been investigated
extensively because they are the building blocks of future integrated circuits.
The mainly features of these quasi-one-dimensional systems are charge conductance
quantization and quantum interference effects, which can be interpreted by the
discrete subbands in the energy dispersion and the Landauer-B$\mathrm{\ddot{u}}$ttiker
formalism \cite{Datta,Ferry}. Recently, much more attention has also been paid to
another degree of freedom of electron, i.e. spin transport in these systems \cite{Zutic,Fabian},
since the prototype of the spin field-effect-transistor (SFET) proposed by Datta and
Das \cite{Datta2}. In the SFET, spin-polarized electrons are injected from a ferromagnetic
lead to a QW, and then the spin is precessed in the wire because the electrons
experience an effective magnetic field induced by spin-orbit coupling (SOC, often referred to
Rashba one \cite{Rashba} or Dresselhaus one \cite{Dresselhaus}) during
transport, and finally the spin is detected by another ferromagnetic lead.
As the SOC presents in the QW, all the spin-degenerated subbands are lifted except for
at the wave vector $k=0$. Therefore, the SOC effect will produce a phase shift of the transmitted electrons
$\Delta\theta=2m^{\ast}\alpha d/\hbar^{2}$ \cite{Mireles}, here $m^{\ast}$ is the effective mass of electron,
$\alpha$ is the SOC strength, and \emph{d} is the transmission distance and $\hbar$ the reduced Plank constant.
More importantly, the Rashba SOC strength has achieved to be tuned by an external electric field in
experiment \cite{Nitta, Grundler}, which provides the potential application of SOC-based spintronics
devices in practice.

In addition to the SOC, the electron spin in QWs is also sensitive to the external magnetic field
(MF) or the proximate ferromagnetic materials \cite{Prinz,Wang,Zhang,Watson}. Differing
from the case of SOC, the spin-degenerated subbands are lifted to Landau levels and each
level is spin split at all the wave vector due to the wave-vector-independent Zeeman effect \cite{Kotlyar}.
However, the Zeeman spin split $\Delta \varepsilon=2\varepsilon_{z}=g^{\ast}\mu_{B}B$ is
very small because of the strong reduction of the effective electron mass
(e.g. $m^{\ast}=0.068 m_{0}$ in GaAs/AlGaAs heterostructure)
and the very low effective Land$\mathrm{\acute{e}}$ factor $g^{\ast}$ in semiconductor materials \cite{Bellucci},
where $\mu_{B}$ is Bohr magneton, $B$ is the MF strength and $m_{0}$ the mass of free electron. Further,
charge and spin transport in a QW is possible only via chiral edge modes when the MF is strong enough
and with strong a robustness against disorders.

More recently, there have also been numerous studies on the spin transport in QWs in
the presence of both SOC(s) and an exteral MF. The interplay of the SOC(s) and MF brings to many new effects.
The first one is the modification of the energy dispersion and conductance. In a
Rashba QW with a perpendicular MF, the Rashba SOC also leads to a Zeeman-like
energy-band split \cite{Wang2,Knobbe,Debald} besides the Zeeman spin split caused by the MF.
However, this type of energy-band split is subband dependent and has a complex dependence
on the MF due to the variance of the expectance of the spin angular momentum operator
along the width of the wire \cite{Pramanik}. However, the Rashba spin precession in quantum-Hall edge
modes is similar to that of Rashba-split QW when the MF is strong \cite{Pala}. In a QW
with SOCs due to different mechanisms such as Rashba and Dresselhaus, as well as
the lateral confining potential, the interaction between these SOCs and the external MF would affect
the transport and optical properties \cite{Zhang3}, and the subbands anticrossings
\cite{Moroz,Zhang2} dramatically. Further, additional subband extrema and energy gap
are also found in a Rashba QW with an in-plane MF \cite{Pershin,Serra}.
The second one is the modification of the transversal spin texture. In the Rashba QW with
an in-plane MF, the spin \emph{x, y} and \emph{z} magnetization along the transversal direction are strongly
dependent on the \emph{k} value and subband index when the Rashba intersubband
coupling is taken into account \cite{Serra}. In the Rashba QW with a perpendicular MF,
additional spin texturing is introduced, and a $\pi/2$ phase shift between the
modulations of the spin density components along the external MF direction and
along the direction of the Rashba-induced effective magnetic field is
observed \cite{Upadhyaya}. However, this effect will disappear when the
MF is stronger than the Rashba-induced effective magnetic field,
which is quite different from the Dresselhaus QW \cite{Gujarathi}. In addition,
other effects including a beating pattern in magnetoresistance \cite{Knobbe,Zhang2}, transverse electron
focusing \cite{Usaj}, resonance spin Hall conductance \cite{Shen} and the suppression
of resonance transmission \cite{Li} have also been found in these systems.

In most works mentioned above, the spin-related transport in the system is
investigated for a spin-polarized injection. Moreover, only the transversal
spin texture has been analyzed, while the longitudinal spin texture has not been
considered seriously thus far. In this paper, using the extended Usuki
transfer-matrix method \cite{Usuki,Akis} combined with the Landauer-B$\mathrm{\ddot{u}}$ttiker
formula, we numerically calculate the spin conductance and the spin-polarized density distributions inside
the Rashba QW in the presence of a perpendicular MF for a spin-unpolarized injection.
Wide energy windows  with three-component spin conductance can be achieved in this system due to the Rashba\_Zeeman effect,
which is quite different from those of the QW with only a perpendicular MF. Moreover, not only the width of these energy windows
but also the magnitude of the spin conductance within these energy windows can be controlled by tuning the Rashba SOC strength,
which can be interpreted respectively by the energy dispersion and the spin-polarized density distributions.
Further study also shows that the spin conductance within these energy windows is robust against the
scattering caused by impurities in the real QW. Thereby, the considered system may find applications in future
spintronics devices.

The rest of this paper is organized as follows. In section 2, the theoretical
model and the spin-resolved Usuki transfer-matrix method are presented. The numerical
results and discussions are shown in section 3. Finally, section 4 concludes the paper.

\section{Theoretical model and the spin-resolved Usuki transfer-matrix method}
The QW studied in present paper is schematically shown in fig. 1, which
is located in a perpendicular MF and sandwiched between two normal
metal leads with the same width $W$ as that of the QW. Only the SOC arising from Rashba mechanism
is considered since its domination in this structure and its strength
can be controlled by an external electrical field. In order to eliminate the
scattering at the interfaces, two buffering regions \cite{Mireles}
(with lengths $L_{1}$ and $L_{3}$, respectively) with adiabatically
variable Rashba SOC and MF strengths are assumed to be situated between the leads
and the middle region (with a length $L_{2}$) with constant ones. Spin-unpolarized
electrons are injected from the left lead and then transported longitudinally
along $x$-axis and confined transversely along $y$-axis and normally along $z$-axis.

Using the Landau gauge, the vector potential is expressed by $\vec{A}=(0, Bx, 0)$.
The $z$-axis is chosen as the spin-quantized axis so that
$|\uparrow\rangle=(1,0)^{\mathrm{T}}$ (here $\mathrm{T}$ means transposition)
represents the spin-up state, $|\downarrow\rangle=(0,1)^{\mathrm{T}}$ denotes the spin-down state,
and the Pauli matrix expressions are
$\sigma_{x}=\left[\matrix{0 & 1\cr 1 & 0} \right]$, $\sigma_{y}=\left[\matrix{0 & -i\cr i & 0} \right]$
and  $\sigma_{z}=\left[\matrix{1 & 0\cr 0 & -1} \right]$. Under these conditions, the single-electron
Schr$\mathrm{\ddot{o}}$dinger equation of the QW at low temperatures reads
\begin{eqnarray}
\{\frac{1}{2m^{\ast }}[p_{x}^{2}+(p_{y}-eBx)^{2}]\sigma_{0}+V(y)\sigma_{0}+\frac{1}{2}g^{\ast}\mu_{B}\sigma_{z}B\nonumber\\
+\frac{\alpha}{\hbar }[\sigma_{x}(p_{y}-eBx)-\sigma_{y}p_{x}]\}{\psi}(x,y)
=E{\psi}(x,y),
\end{eqnarray}
where $\sigma_{0}$ is the unit $(2\times2)$ matrix, $V(y)$ is the transversal confining potential,
$E$ and ${\psi}(x,y)=\left[\matrix{\psi^{\uparrow}(x,y)\cr \psi^{\downarrow}(x,y)}\right]$
are the energy and spin-dependent wavefunction of electron, respectively.
As well known, the analytic solution of this equation is very hard to be obtained.
However, its numerical solution is easy to be achieved by discretizing it
on a rectangular grid, with the indexes $l$ and $m$ respectively representing the sites along the
$x$- and $y$-axis. Under the tight-binding approximation, Eq. (1) can be written as
\begin{eqnarray}
(E\textbf{I}-\textbf{H}_{l}){\psi}_{l,m}-\textbf{H}_{l,l+1}{\psi}_{l+1,m}-\textbf{H}_{l,l-1}{\psi}_{l-1,m}=0,
\end{eqnarray}
where $\textbf{I}$ is the unit $(2M\times 2M)$ matrix, here $M$ is the lattice number of each column cell.
$\textbf{H}_{l}=\left[\matrix{\textbf{H}_{l}^{\uparrow\uparrow} & \textbf{H}_{l}^{\uparrow\downarrow}\cr \textbf{H}_{l}^{\downarrow\uparrow} & \textbf{H}_{l}^{\downarrow\downarrow}}\right]$
is the Hamiltonian of the $l$th isolated column cell in both spatial and spin spaces.
$\textbf{H}_{l,l+1}=\left[\matrix{\textbf{H}_{l,l+1}^{\uparrow\uparrow} & \textbf{H}_{l,l+1}^{\uparrow\downarrow}\cr \textbf{H}_{l,l+1}^{\downarrow\uparrow} & \textbf{H}_{l,l+1}^{\downarrow\downarrow}}\right]$
is the intercell Hamiltonian between the $l$th column cell and the $(l+1)$th column cell, and $\textbf{H}_{l,l-1}=(\textbf{H}_{l,l+1})^\dag$. The explicit expression for
each spin-resolved term is
\begin{eqnarray}
\textbf{H}_{l}^{\uparrow\uparrow/\downarrow\downarrow}=\left[\matrix{4.0t+V_{l,1}\mp\varepsilon_{z} & -t & 0 & \cdots & 0 \cr
                                                            -t & 4.0t+V_{l,2}\mp\varepsilon_{z} & -t & \ddots & \vdots \cr
                                                            0 & -t & \ddots & \ddots & 0 \cr
                                                            \vdots & \ddots & \ddots & \ddots & -t \cr
                                                            0 & \cdots & 0 & -t & 4.0t+V_{l,M}\mp\varepsilon_{z} }\right],
\end{eqnarray}
\begin{eqnarray}
\textbf{H}_{l}^{\uparrow\downarrow}=\textbf{H}_{l}^{\downarrow\uparrow}=\left[\matrix{0 & it_{so} & 0 & \cdots & 0 \cr
                                                            -it_{so} & 0 & it_{so} & \ddots & \vdots \cr
                                                            0 & -it_{so} & \ddots & \ddots & 0 \cr
                                                            \vdots & \ddots & \ddots & \ddots & it_{so} \cr
                                                            0 & \cdots & 0 & -it_{so} & 0 }\right],
\end{eqnarray}
\begin{eqnarray}
\textbf{H}_{l,l+1}^{\uparrow\uparrow}=\textbf{H}_{l,l+1}^{\downarrow\downarrow}=\left[\matrix{-e^{(-i\frac{\hbar\omega_{c}}{2t})} & 0 & 0 & \cdots & 0 \cr
                                                                    0 & -e^{(-i\frac{\hbar\omega_{c}}{t})} & 0 & \ddots & \vdots \cr
                                                                    0 & 0 & \ddots & \ddots & 0 \cr
                                                                    \vdots & \ddots & \ddots & \ddots & 0 \cr
                                                                    0 & \cdots & 0 & 0 & -e^{(-i\frac{\hbar\omega_{c}}{2t}M)} }\right],
\end{eqnarray}
\begin{eqnarray}
\textbf{H}_{l,l+1}^{\uparrow\downarrow}=-\textbf{H}_{l,l+1}^{\downarrow\uparrow}=t_{so}\textbf{H}_{l,l+1}^{\uparrow\uparrow/\downarrow\downarrow},
\end{eqnarray}
in which $t=\hbar^{2}/2m^{\ast}a^{2}$ is the hopping energy with the lattice constant
$a$. $\omega_{c}=eB/m^{\ast}c$ is the cyclotron frequency and $t_{so}=\frac{\alpha}{2a}$.

Both the propagating and evanescent modes can be obtained by combining the Bloch's
theorem with the eigenvalue problem for the transfer-matrix form of Eq. (2)
\begin{eqnarray}
\left[\matrix{0 & \textbf{I} \cr -\textbf{H}_{l,l+1}^{-1}\textbf{H}_{l,l-1} & \textbf{H}_{l,l+1}^{-1}(E\textbf{I}-\textbf{H}_{l}) }\right]
\left[\matrix{{\psi}_{l-1,m}\cr {\psi}_{l,m}} \right]=\lambda \left[\matrix{{\psi}_{l-1,m}\cr {\psi}_{l,m}} \right],
\end{eqnarray}
in which $\lambda$ is a phase factor of a plane wave along the $x$ axis. As a result,
this equation has $4M$ eigenvalues $\lambda_{j}$ and eigenvectors
$\textbf{u}_{j}$, which can be classified into $2M$ right-moving wave [$\lambda_{j}(+)$, $\textbf{u}_{j}(+)$] and
$2M$ left-moving waves [$\lambda_{j}(-)$, $\textbf{u}_{j}(-)$] \cite{Khomyakov}. For the scattering problem of
the wave function in the considered system, the spin-resolved matrices $\textbf{t}$ and $\textbf{r}$ of
the transmission and reflection waves are obtained by
\begin{eqnarray}
\left[\matrix{\textbf{t} \cr \textbf{0} }\right]=\textbf{T}_{0}^{-1}\textbf{T}_{L}\cdots \textbf{T}_{l}\cdots\textbf{T}_{0}\left[\matrix{\textbf{I} \cr \textbf{r} }\right],
\end{eqnarray}
in which $L$ is the lattice number along the $y$-axis. $\textbf{I}$ means the modes injected from the left lead with unit amplitude.
These $(4M\times 4M)$ transfer matrices are given by
\begin{eqnarray}
\textbf{T}_{0}=\left[\matrix{\textbf{U}(+) & \textbf{U}(-) \cr \textbf{U}(+){\lambda}(+) & \textbf{U}(-){\lambda}(-) }\right],
\end{eqnarray}
with $\textbf{U}(\pm)=[\textbf{u}_{1}(\pm),\cdots \textbf{u}_{j}(\pm),\cdots,\textbf{u}_{2M}(\pm)]$ and ${\lambda}(\pm)=\mathrm{diag}[\lambda_{1}(\pm),\cdots \lambda_{j}(\pm),\cdots,\lambda_{2M}(\pm)]$.
\begin{eqnarray}
\textbf{T}_{l}=\left[\matrix{\textbf{T}_{l11} & \textbf{T}_{l12} \cr \textbf{T}_{l21} & \textbf{T}_{l22} }\right]\nonumber\\
=\left[\matrix{\textbf{0} & \textbf{I} \cr -\textbf{H}_{l,l+1}^{-1}\textbf{H}_{l,l-1} & \textbf{H}_{l,l+1}^{-1}(E\textbf{I}-\textbf{H}_{l}) }\right]~~\mathrm{for}~~1\leq l\leq L.
\end{eqnarray}

According to the spin-resolved transmission and reflection matrices obtained from Eq. (8), one can evaluate the spin-resolved transmission
and reflection conductances by using the Landauer-B$\mathrm{\ddot{u}}$ttiker generalized to include the spin degree of freedom
\begin{eqnarray}
\textbf{G}=\left[\matrix{G^{\uparrow\uparrow} & G^{\uparrow\downarrow} \cr G^{\downarrow\uparrow} & G^{\downarrow\downarrow} }\right]
          =\frac{e^{2}}{h}\sum\limits_{\mu,\nu=1}^{M}
          \left[\matrix{|\textbf{t}_{\nu\mu}^{\uparrow\uparrow}|^{2} & |\textbf{t}_{\nu\mu}^{\uparrow\downarrow}|^{2} \cr |\textbf{t}_{\nu\mu}^{\downarrow\uparrow}|^{2} & |\textbf{t}_{\nu\mu}^{\downarrow\downarrow}|^{2} }\right],
\end{eqnarray}
\begin{eqnarray}
\textbf{R}=\left[\matrix{R^{\uparrow\uparrow} & R^{\uparrow\downarrow} \cr R^{\downarrow\uparrow} & R^{\downarrow\downarrow} }\right]
          =\frac{e^{2}}{h}\sum\limits_{\mu,\nu=1}^{M}
          \left[\matrix{|\textbf{r}_{\nu\mu}^{\uparrow\uparrow}|^{2} & |\textbf{r}_{\nu\mu}^{\uparrow\downarrow}|^{2} \cr |\textbf{r}_{\nu\mu}^{\downarrow\uparrow}|^{2} & |\textbf{r}_{\nu\mu}^{\downarrow\downarrow}|^{2} }\right],
\end{eqnarray}
where $\textbf{t}_{\nu\mu}^{\sigma'\sigma}$ $(\textbf{r}_{\nu\mu}^{\sigma'\sigma})$ means
the spin-dependent transmission (reflection) coefficient from the incident mode $\mu$
with spin $\sigma$ to the out-going mode $\nu$ with spin $\sigma'$ in the right (left) lead.

In general,  Eq. (8) is extremely unstable due to the exponentially growing and decaying contributions of
the evanescent modes when the product of transfer matrices is taken. However, this
unstability can be overcomed by the following iteration technique proposed by Usuki \cite{Usuki}:
\begin{eqnarray}
\left[\matrix{\textbf{C}_{1}^{l+1} & \textbf{C}_{2}^{l+1} \cr \textbf{0} & \textbf{I} }\right]=
\textbf{T}_{l}\left[\matrix{\textbf{C}_{1}^{l} & \textbf{C}_{2}^{l} \cr \textbf{0} & \textbf{I} }\right]\textbf{P}_{l}~~\mathrm{for}~~0\leq l\leq L+1,
\end{eqnarray}
with
\begin{eqnarray}
\textbf{T}_{L+1}=\left[\matrix{\textbf{0} & [\textbf{U}(+){\lambda}(+)]^{-1} \cr \textbf{I} & -\textbf{U}(+)[\textbf{U}(+){\lambda}(+)]^{-1} }\right],
\end{eqnarray}
\begin{eqnarray}
\textbf{P}_{l}=\left[\matrix{\textbf{1} & \textbf{0} \cr \textbf{P}_{l1} & \textbf{P}_{l2} }\right],
\end{eqnarray}
\begin{eqnarray}
\textbf{P}_{l2}=(\textbf{T}_{l21}\textbf{C}_{2}^{l}+\textbf{T}_{l22})^{-1},
\end{eqnarray}
and
\begin{eqnarray}
\textbf{P}_{l1}=-\textbf{P}_{l2}\textbf{T}_{l21}\textbf{C}_{1}^{l}.
\end{eqnarray}
The iteration continues from $l=0$ to $L+1$ under an initial condition $\textbf{C}_{1}^{0}=\textbf{I}$ and $\textbf{C}_{2}^{0}=\textbf{0}$.
Finally, the spin-resolved transmission matrix $\textbf{t}=\textbf{C}_{1}^{L+2}$ can be obtained in the last step of the iteration.
Similarly, the spin-resolved reflection matrix $\textbf{r}=\textbf{D}_{1}^{L+2}$ is given by iteration
\begin{eqnarray}
\left(\matrix{\textbf{D}_{1}^{l+1} & \textbf{D}_{2}^{l+1} }\right)=
\left(\matrix{\textbf{D}_{1}^{l} & \textbf{D}_{2}^{l} }\right)\textbf{P}_{l}~~\mathrm{for}~~0\leq l\leq L+1,
\end{eqnarray}
with an initial condition $\textbf{D}_{1}^{0}=\textbf{0}$ and $\textbf{D}_{2}^{0}=\textbf{I}$.

Besides the spin-resolved transmission and reflection conductances, the spin-resolved
electron wave functions inside the quantum wire can also be reconstructed by
using the same matrices $\textbf{P}_{l1}$ and $\textbf{P}_{l2}$ calculated above.
However, the procedure of the iteration is going from the final column cell
(right) to the initial column cell (left) of the considered system \cite{Akis},
which is inverse to that of the conductance calculation.
The explicit iteration equation is given by
\begin{eqnarray}
{\phi}_{l-1,m}^{(j)}=\textbf{P}_{(l-1)1}+\textbf{P}_{(l-1)2}{\phi}_{l,m}^{(j)}~~\mathrm{for}~~L+1\geq l> 1,
\end{eqnarray}
with the initial condition is defined as ${\phi}_{(L+1),m}^{(j)}=\textbf{P}_{(L+1)1}$.
Now the amplitude matrix of the spin-resolved electron wave function at each column cell can be achieved during the iteration process
\begin{eqnarray}
\textbf{a}_{l,m,j}=\left[\matrix{\textbf{a}_{l,m,j}^{\uparrow\uparrow} & \textbf{a}_{l,m,j}^{\uparrow\downarrow} \cr \textbf{a}_{l,m,j}^{\downarrow\uparrow} & \textbf{a}_{l,m,j}^{\downarrow\downarrow} }\right]={\phi}^{(j)}_{l,m}.
\end{eqnarray}
where $j$ denotes the propagating mode in the injected lead.

\section{Numerical results and discussions}

In the following numerical calculations, all the energies are normalized by the hopping energy $t$ ($t=1$) and all
the lengths are normalized by the lattice constant $a$ ($a=1$).
The structural parameters of the considered system are taken as $W=M+1=20$ and $L_{1}=L_{2}=L_{3}=\frac{L-1}{3}=40$.
The Rashba SOC and MF strengths as a function of the index $l$ are given as
\begin{eqnarray}
  t_{so}(l)=\left\{
  \begin{array}{ll}
    t_{so}\sin\frac{(l-1)\pi}{80}, & 1\leq l\leq 41 \\
    t_{so}, & 42\leq l\leq 80 \\
    t_{so}\sin\frac{(121-l)\pi}{80}, & 81\leq l\leq 121
  \end{array}
\right.
\end{eqnarray}
and
\begin{eqnarray}
  \hbar\omega_{c}(l)=\left\{
  \begin{array}{ll}
    \hbar\omega_{c}\sin\frac{(l-1)\pi}{80}, & 1\leq l\leq 41 \\
    \hbar\omega_{c}, & 42\leq l\leq 80 \\
    \hbar\omega_{c}\sin\frac{(121-l)\pi}{80}, & 81\leq l\leq 121
  \end{array}
\right.
\end{eqnarray}
\begin{eqnarray}
  \varepsilon_{z}(l)=\left\{
  \begin{array}{ll}
    \varepsilon_{z}\sin\frac{(l-1)\pi}{80}, & 1\leq l\leq 41 \\
    \varepsilon_{z}, & 42\leq l\leq 80 \\
    \varepsilon_{z}\sin\frac{(121-l)\pi}{80}, & 81\leq l\leq 121
  \end{array}
\right.
\end{eqnarray}
where $\hbar\omega_{c}=0.2$ and $\varepsilon_{z}=0.02$. In addition, the hard-wall
confining potential approximation is adopted to the transversal confining potential,
that is, $V_{l,m}=0$ for $1\leq m \leq M$ and $\infty$ otherwise.
The transmission charge conductance and the transmission spin conductance vector are defined as
\begin{eqnarray}
G^{e}=G^{\uparrow\uparrow}+G^{\uparrow\downarrow}+G^{\downarrow\downarrow}+G^{\downarrow\uparrow},
\end{eqnarray}
and
\begin{eqnarray}
\textbf{G}^{\textbf{S}}=(G^{S_x}, G^{S_y}, G^{S_z}),
\end{eqnarray}
respectively. Each component of the transmission spin conductance vector in Eq. (25) can be calculated by \cite{Nikolic0}
\begin{eqnarray}
G^{S_x}=\frac{e}{4\pi}\sum\limits_{\mu,\nu=1}^{M}\mathrm{Re}[\textbf{t}_{\nu\mu}^{\uparrow\uparrow}(\textbf{t}_{\nu\mu}^{\downarrow\uparrow})^{\ast}
+\textbf{t}_{\nu\mu}^{\uparrow\downarrow}(\textbf{t}_{\nu\mu}^{\downarrow\downarrow})^{\ast}],
\end{eqnarray}
\begin{eqnarray}
G^{S_y}=\frac{e}{4\pi}\sum\limits_{\mu,\nu=1}^{M}\mathrm{Im}[(\textbf{t}_{\nu\mu}^{\uparrow\uparrow})^{\ast}\textbf{t}_{\nu\mu}^{\downarrow\uparrow}
+(\textbf{t}_{\nu\mu}^{\uparrow\downarrow})^{\ast}\textbf{t}_{\nu\mu}^{\downarrow\downarrow}],
\end{eqnarray}
and
\begin{eqnarray}
G^{S_z}=\frac{e}{4\pi}\frac{G^{\uparrow\uparrow}+G^{\uparrow\downarrow}-G^{\downarrow\downarrow}-G^{\downarrow\uparrow}}{e^2/h}.
\end{eqnarray}
Similarly, the reflection charge conductance is defined as
\begin{eqnarray}
R^{e}=R^{\uparrow\uparrow}+R^{\uparrow\downarrow}+R^{\downarrow\downarrow}+R^{\downarrow\uparrow}.
\end{eqnarray}
And the local spin-polarized density vector at each column cell are given by
\begin{eqnarray}
{\rho}_{S_{x}}(l,m)=\sum\limits_{j=1}^{M}\mathrm{Re}[\textbf{a}_{l,m,j}^{\uparrow\uparrow}(\textbf{a}_{l,m,j}^{\downarrow\uparrow})^{\ast}
+\textbf{a}_{l,m,j}^{\uparrow\downarrow}(\textbf{a}_{l,m,j}^{\downarrow\downarrow})^{\ast}],
\end{eqnarray}
\begin{eqnarray}
{\rho}_{S_{y}}(l,m)=\sum\limits_{j=1}^{M}\mathrm{Im}[(\textbf{a}_{l,m,j}^{\uparrow\uparrow})^{\ast}\textbf{a}_{l,m,j}^{\downarrow\uparrow}
+(\textbf{a}_{l,m,j}^{\uparrow\downarrow})^{\ast}\textbf{a}_{l,m,j}^{\downarrow\downarrow}],
\end{eqnarray}
and
\begin{eqnarray}
{\rho}_{S_{z}}(l,m)=\sum\limits_{j=1}^{M}(|\textbf{a}_{l,m,j}^{\uparrow\uparrow}|^{2}+|\textbf{a}_{l,m,j}^{\uparrow\downarrow}|^{2}-
|\textbf{a}_{l,m,j}^{\downarrow\downarrow}|^{2}-|\textbf{a}_{l,m,j}^{\downarrow\uparrow}|^{2}).
\end{eqnarray}

Figure 2(a) shows the transmission (the black solid line) and reflection
(the red dashed line) charge conductances as a function of the electron
energy for the QW with only a perpendicular MF. Perfect step-shaped
structures are found in both the charge and reflection charge conductances
because of the insertion of the two buffering regions between the leads
and the middle region with constant MF strength, which suppresses the
scattering due to the mismatch of the energy dispersions in the middle wire
and leads. As the MF presents in the QW, Landau energy subbands with Zeeman
spin split are formed in the energy dispersion, as shown in fig. 2(b), which
determines the transmission charge conductance of the whole system. Therefore,
steps with the magnitude of odd numbers of conductance quantization ($e^{2}/h$)
emerge in the charge conductance spectra. In addition, the total magnitude
of the transmission and reflection charge conductances
of the whole system (the blue dotted line) exactly equals that of the charge
conductance contributed from the propagating modes in the injected lead.
Figure 2(c) plots the transmission charge conductance as a function of
the electron energy and Rashba SOC strength for the
QW with both Rashba SOC and a perpendicular MF. Similar to the case in fig. 2(a),
ideal quantized conductance steps are also found in the transmission charge conductance
spectra, as shown in the top inset of fig. 2(c). This transport behavior
can be elucidated by the energy dispersion in fig. 2(d), where the strength of
Rashba SOC is set at $t_{so}=0.05$ (see the yellow horizontal line).
However, the width of each charge conductance steps
can be tuned by varying the Rashba SOC strength, i.e. the transmission
charge conductance at a certain energy can hop from a step to another.
A concrete example is shown in the right inset of fig. 2(c),
where the electron energy is taken as $E=0.2$ (see the yellow vertical line). The magnitude of
the charge conductance transits from $2e^{2}/h$ to $3e^{2}/h$ when the Rashba SOC
strength is increase to $t_{so}=0.124$.

Figure 3(a) demonstrates the transmission spin conductance as a function of the electron energy
for the QW with a perpendicular MF. Only the $z$-component transmission spin conductance is
achieved when the electron energy is located within the two thresholds of each pair
of Landau energy subbands, namely, energy windows with non-vanishing spin conductance
can be obtained. Further, all these energy windows are identical and with the same
width and magnitude. However, for the QW with both Rashba SOC and a perpendicular MF,
the transmission spin conductance shows more complicate behaviors. First, all
three components of the transmission spin conductance, as shown respectively in figs.
3(b), 3(c) and 3(d), are generated when the electron energy is situated within the
energy windows caused by Rashba\_Zeeman spin split. Second, the spin conductance within the energy
windows has both subband index and energy dependence, as shown in the upper insets of figs.
3(b), 3(c) and 3(d), in which the Rashba SOC strength is taken as $t_{so}=0.05$
(as indicated by the yellow horizontal lines). Third, the spin conductance within the
energy windows can be manipulated by varying the Rashba SOC strength, as shown in
the left insets of figs. 3(b), 3(c) and 3(d), where the electron energy is set at
$E=0.1$ (as indicated by the yellow vertical lines). This effect is attributed to
interaction between the effective magnetic field induced by the
Rashba SOC and the Zeeman spin split, resulting in the variance of the spin
conductance. Final, the widths of these energy windows are sensitive to
the subband index and Rashba SOC strength. The width of each energy widow
is enlarged with the increase of Rashba SOC strength and can be distinguished from
each other for the weak Rashba SOC strength. However, as the Rashba SOC strength
is increased further, each neighboring two energy window will overlap,
leading to the generation of spin conductance at all electron energies.

In order to understand the spin conductance of the QW in the presence of both
Rashba SOC and a perpendicular MF obtained in figs. 3, the $z$-component spin-polarized density
distributions inside the QW at different electron energies and Rashba SOC
strengths are plotted in fig. (4). The explicit parameters in each panel are
(a) $E_{1}=0.078$ and $t_{so1}=0.05$, (b) $E_{1}=0.078$ and $t_{so2}=0.1$,
(c) $E_{2}=0.1$ and $t_{so1}=0.05$, (d) $E_{2}=0.1$ and $t_{so2}=0.1$.
Highly spin-polarized density island with negative sign is formed in the buffering
region to the left, originating from the interaction between the bound state and the
external MF and the effective MF induced by Rashba SOC \cite{Xiao}. However, spin-polarized
density ribbon with positive sign is observed in the lower edge of the middle region of the QW,
which is attributed to the interaction between the chiral edge state caused by the external MF and the
Rashba SOC. Therefore, the magnitude of the spin-polarized density ribbon can be tuned
by varying the electron energy within the energy windows and the Rashba SOC strength,
which consists with the spin conductance properties of the whole system shown in fig. 3. In addition,
the spin-polarized density distributions of the $x$- and $y$-component inside the QW
display the same features as those of the $z$-component so that not be presented here.

The above calculations assume a perfectly QW, where there is no elastic or inelastic scattering.
However, in a realistic QW, there will be many impurities in the sample. Consequently, the effect of
disorder on the spin-dependent transport in the real QW should be considered in practical applications.
The effects of impurities can be introduced by fluctuation of the diagonal terms
of Eq. (3), which are distributed randomly within a range of
width $w$: $\mathrm{diag}(\textbf{H}_{l}^{\uparrow\uparrow/\downarrow\downarrow})=
\mathrm{diag}(\textbf{H}_{l}^{\uparrow\uparrow/\downarrow\downarrow})+w_{lm}$, here $-w/2<w_{lm}<w/2$.
Figure 5 shows the average transmission charge and spin conductances as a function of the electron
energy for weak ($w=0.2$, the red dashed lines) and strong ($w=0.4$, the blue dotted
lines) disorder strengths. Number of the real samples taken for calculating the average
values is $1000$. In contrast to the transmission charge conductance of the perfect QW
with the same parameters (the black solid line), as shown in fig. 5(a), the
step-like structures disappear in the average charge conductance
as disorder presents in the QW. Moreover, dip-like structures emerge at the average charge
conductance and their positions just around the ends of the charge conductance steps,
resulting from the interplay of the disorder-induced bound states and
the continued states in the leads. However, the amplitude of the
transmission charge conductance does not drop much even when the strong
disorder is presented in the QW. Similar to the average charge
conductance, the average spin conductance within the energy windows is also
destroyed by the disorder, as shown in figs. 5(b), 5(c) and 5(d).
However, the magnitude of the average spin conductance within the energy windows is still
large even in the presence of a strong disorder, especially for the lower energy windows.
These effects may be attributed to the edge state caused by the MF, as shown in fig. 4,
which is immune from the scatter of the impurities. Therefore,
both the charge and spin conductances can survive in the disordered QW.

\section{Conclusion}

In conclusion, spin-dependent transport properties of a QW in the presence of both Rashba SOC and a perpendicular
MF for a spin-unpolarized injection is studied by using the extended Usuki transfer-matrix method
together with the Landauer-B$\mathrm{\ddot{u}}$ttiker formalism. A spin-polarized
current of three components can be generated in the output lead when the electron energy lies in the
energy gaps induced by the Rashba\_Zeeman effects and its magnitude can be controlled
by varying the Rashba strength. The mechanism of the generated spin-polarized current
in the output lead is clarified by analyzing both the transversal and longitudinal
spin-polarized density inside the QW. Further study also shows that the spin-polarized current
can survive even in the presence of a strong disorder. Although an external MF is needed to
achieve the spin-polarized current, it is only used to break the time inversion
symmetry. The width of spin-filtering energy gaps and the magnitude of the
spin-polarized current is also manipulated by varying the Rashba SOC strength, indicating that
the considered system may has a potential application in designing
a spin filter device.

\section*{Acknowledgments}

This work was supported by the National Natural Science Foundation
of China (Grant Nos. 11264019, 11274108 and 11304010) and by the development project
on the young and middle-aged teachers in the colleges and universities of Jiangxi.

\newpage
\section*{Figure captions}

~~~

Figure 1. (Color online) Schematic diagrams of a QW with both
Rashba SOC and a perpendicular MF, connected to two semi-infinite normal metal leads.
The three regions of the QW have the same width $W$
but different lengths $L_{1}$,  $L_{2}$ and  $L_{3}$.

~~~

Figure 2. (Color online) (a) The transmission (the black solid line) and reflection
(the red dashed line) charge conductances as a function of the electron energy for the QW with only
a perpendicular MF. The blue dotted line represents the total of the transmission and reflection charge
conductances. (c) The transmission charge conductance as a function of the electron energy and Rashba
SOC strength for the QW with both Rashba SOC and a perpendicular MF. (b) and (d) are the corresponding
energy dispersions for the cases in (a) and (c), respectively. The strength of the
Rashba SOC in (d) is $t_{so}=0.05$.

~~~

Figure 3. (Color online) (a) The transmission spin conductance vector as a function of
the electron energy for the QW with only a perpendicular MF. (b-d) The transmission spin
conductance vector as a function of the electron energy and Rashba
SOC strength for the QW with both Rashba SOC and a perpendicular MF. The Rashba SOC strength
in the upper insets is taken as $t_{so}=0.05$ and the electron energy in the right
inset is $E=0.1$.

~~~

Figure 4. (Color online) The $z$-component spin-polarized density distributions
inside the QW  with both Rashba SOC and a perpendicular MF. The
electron energy and Rashba SOC strength in each panel are taken as (a) $E_1=0.078$ and $t_{so1}=0.05$,
(b) $E_1=0.078$ and $t_{so2}=0.1$, (c) $E_2=0.1$ and $t_{so1}=0.05$, and
(d) $E_2=0.1$ and $t_{so2}=0.1$.

~~~

Figure 5. (Color online) The average transmission charge (a) and spin (b-d) conductances
as a function of the electron energy for the disordered QW with both Rashba SOC and a perpendicular MF.
The disorder strengths are taken as $w=0.2$ (the red dashed lines) and $0.4$ (the blue dotted lines).
The Rashba SOC strength is $t_{so}=0.05$. Number of samples taken for calculating average value is 1000.
These results can be compared with the results for the case of the perfect QW with same parameters
(the black solid lines).

\newpage

\begin{figure}
\center
\includegraphics[width=5.0in]{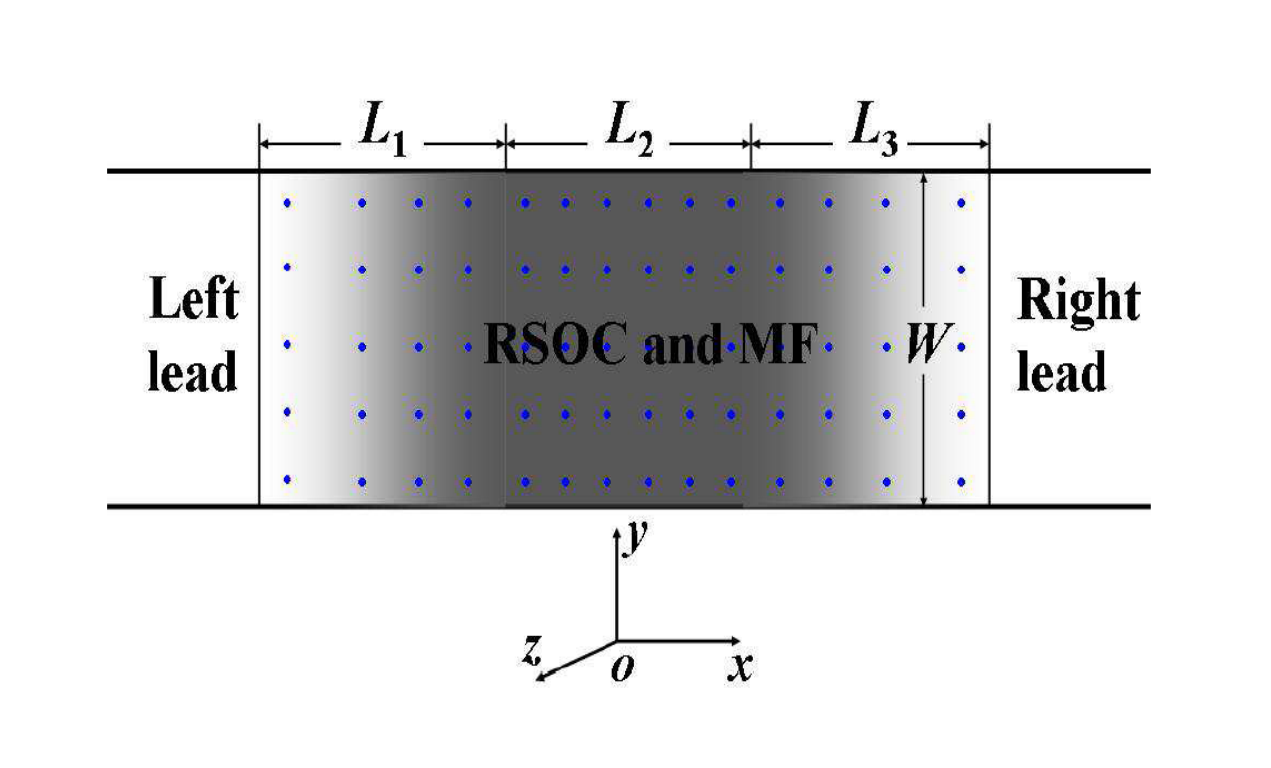}
\center{Figure 1}
\end{figure}

\begin{figure}
\center
\includegraphics[width=5.0in]{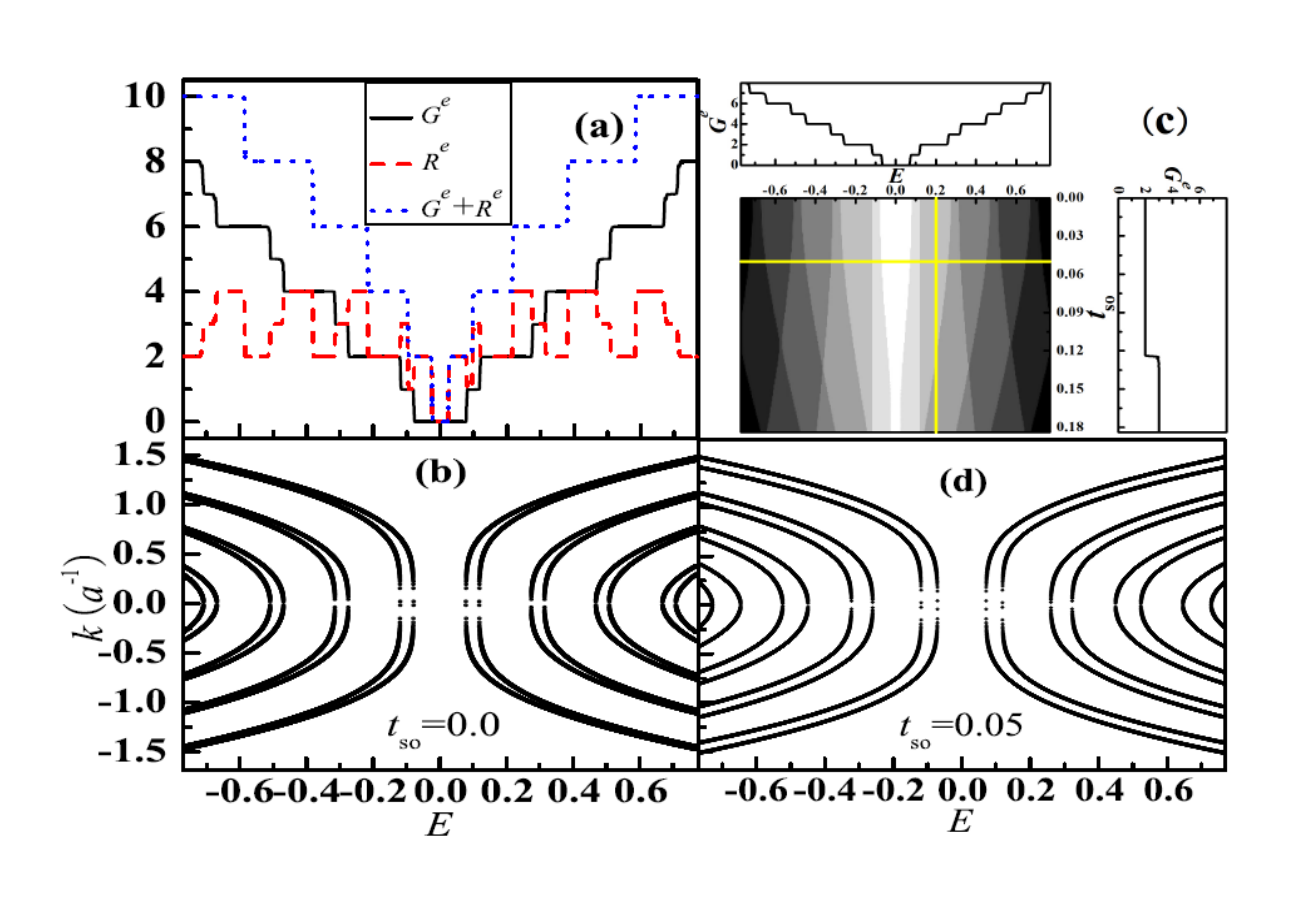}
\center{Figure 2}
\end{figure}

\begin{figure}
\center
\includegraphics[width=5.0in]{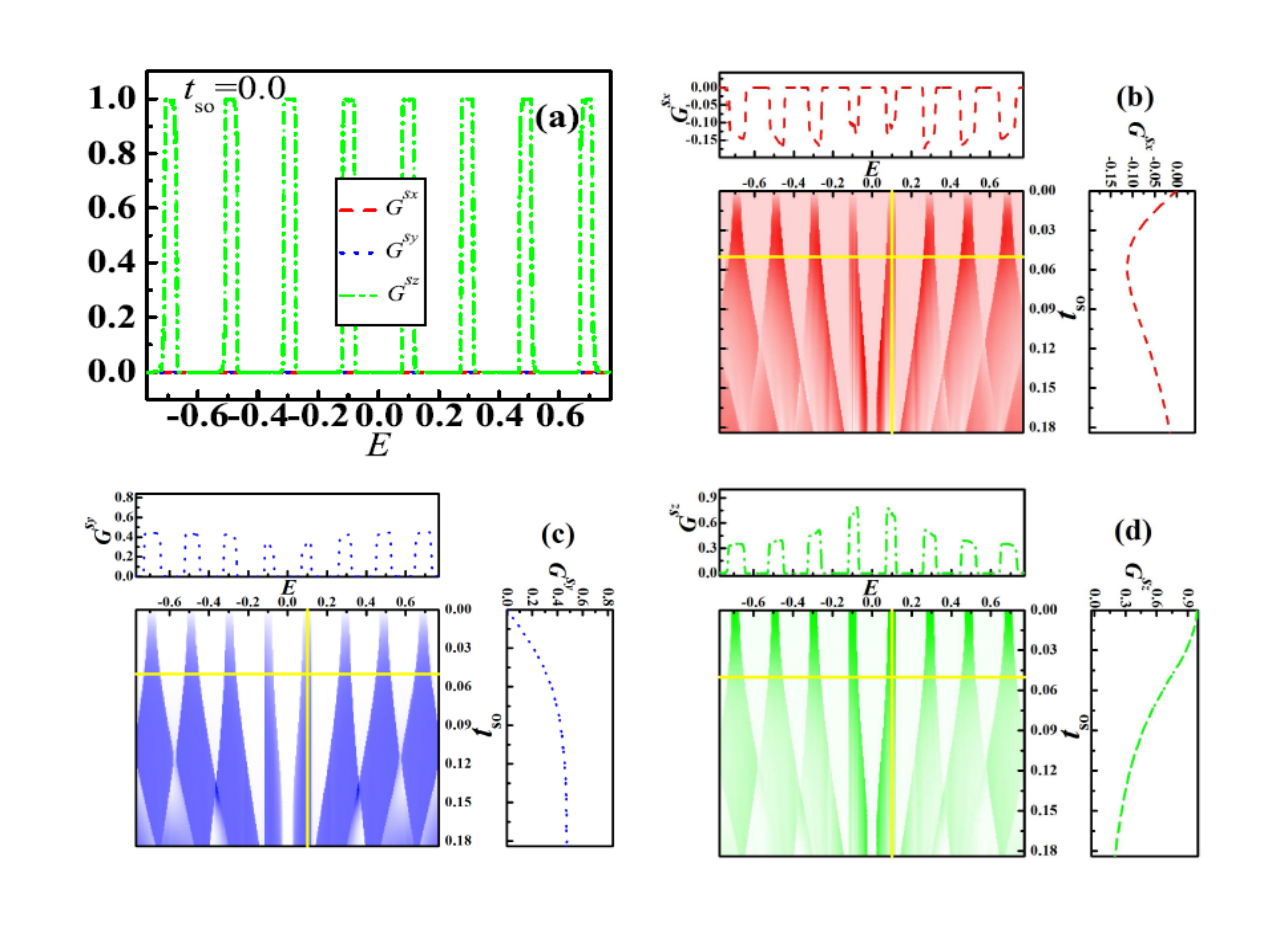}
\center{Figure 3}
\end{figure}

\begin{figure}
\center
\includegraphics[width=5.0in]{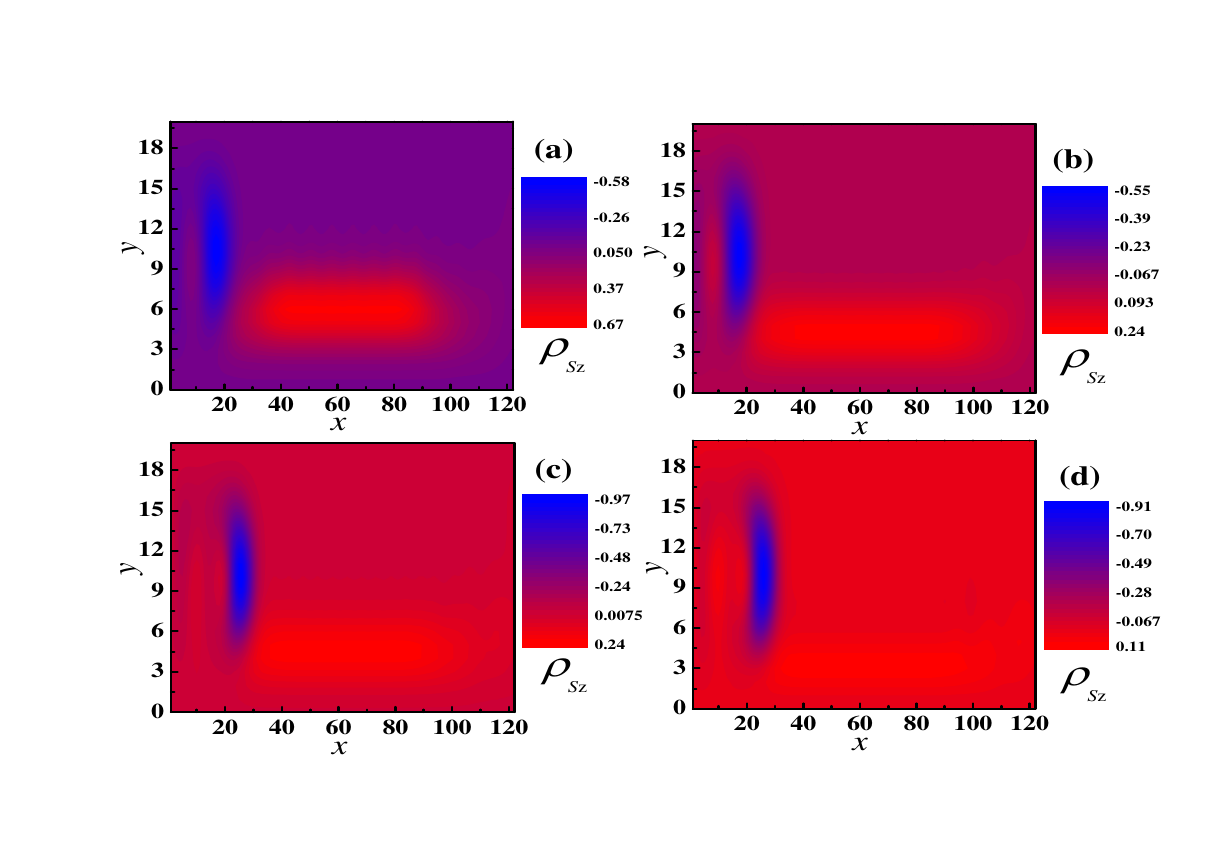}
\center{Figure 4}
\end{figure}

\begin{figure}
\center
\includegraphics[width=5.0in]{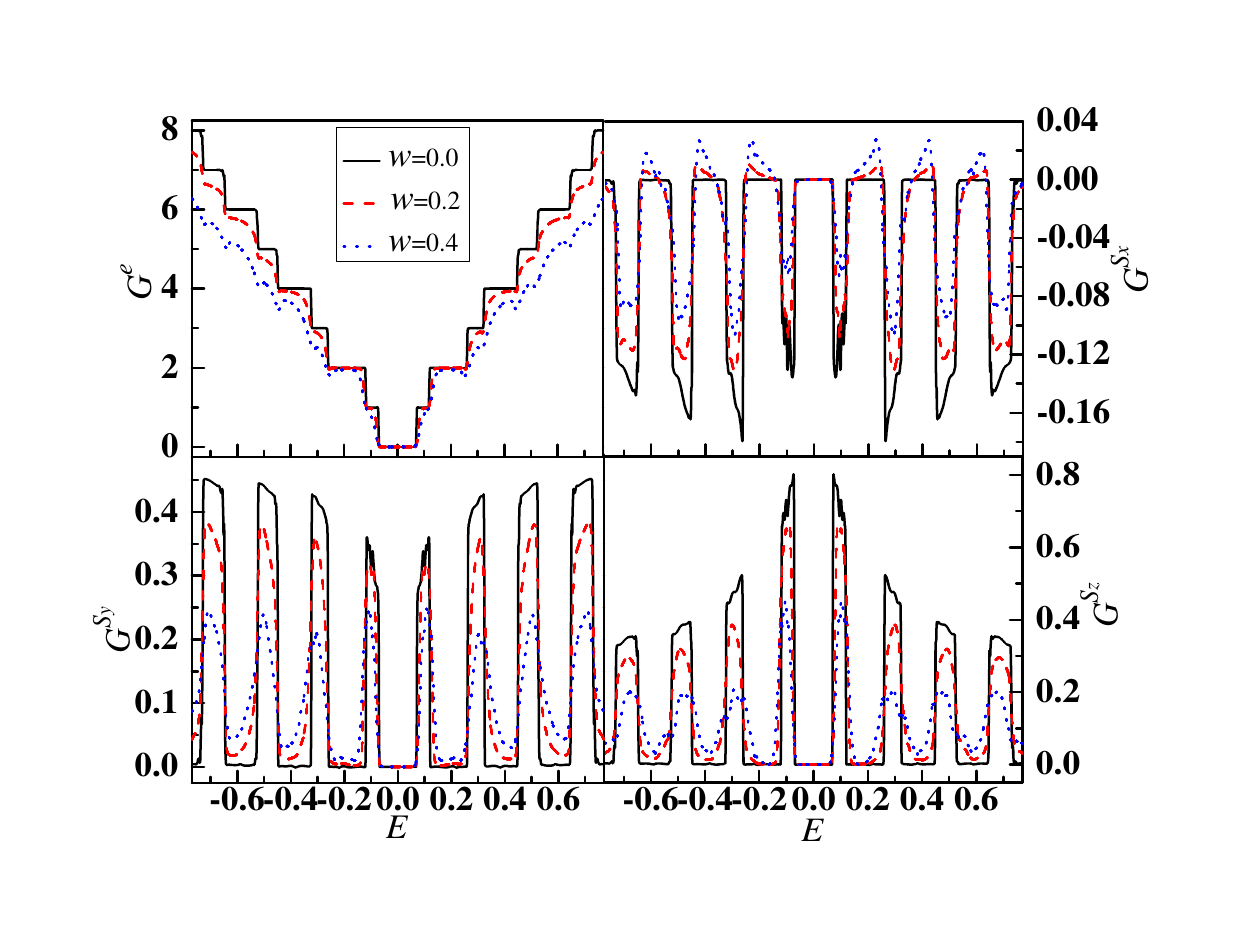}
\center{Figure 5}
\end{figure}

\end{document}